\documentclass[twocolumn,showpacs,preprintnumbers,amsmath,amssymb]{revtex4}

\usepackage{graphicx}
\usepackage{dcolumn}
\usepackage{bm}

\begin{document}

\title{Asymptotic behavior of the entropy of chains placed on stripes.}

\author{W.G. Dantas}
 \email{wgd@gibbs.if.usp.br}
\author{M.J. de Oliveira}%
 \email{oliveira@if.usp.br}
\affiliation{
Universidade de S\~{a}o Paulo,
Caixa Postal 66318
05315-970 S\~{a}o Paulo, S\~{a}o Paulo, Brazil}

\author{J.F. Stilck}
 \email{jstilck@if.uff.br}
\affiliation{Instituto de F\'{\i}sica, Universidade Federal Fluminense,
24210-340, Niter\'oi, RJ, Brazil}

\date{\today}

\begin{abstract}
By using the transfer matrix approach, we investigate the asymptotic behavior
of the entropy of flexible chains with $M$ monomers each placed on
stripes. In the limit of high density of monomers, 
we study the behavior of the entropy as a function of the density
of monomers and the width of the stripe, inspired by recent analytical
studies of this problem for the particular case of dimers
($M=2$). We obtain the entropy in the asymptotic regime of high
densities for chains 
with $M=2,..,9$ monomers, as well as for the special case of polymers, 
where $M\rightarrow\infty$, and find that the results show a regular
behavior similar to the one found analytically for dimers. We also
verify that in the low-density 
limit the mean-field expression for the entropy is followed by the
results from our transfer matrix calculations.
\end{abstract}


\maketitle

\section{Introduction}
\label{intro}

In the thirties, the dimer model was introduced to mimic the adsorption
of diatomic molecules on a crystal  surface \cite{fowler37}. Later,
this model  
was applied in the study of many other physical systems as ferroelectric
and ferromagnetic materials \cite{kast2,fis2,fan,sn74,nagle}. The dimers
can be modeled as chains 
with two basic units called  monomers, occupying first neighbor sites
of a lattice. A central question in the study of this
model is to enumerate the number of ways to place $p$ dimers on
lattice with $N$ sites, 
such that the density of monomers is given by $\rho=2p/N$. The special
case of full occupancy, where $\rho=1$, 
was exactly solved for planar lattices, using a technique based on
pfaffians \cite{fis,kast,temp}. 
However, the more general case $\rho<1$, the so called monomer-dimer
problem, is still an open question. Recently, 
an analytic solution was obtained for the case where there is a single
vacancy at  
a certain site on the boundary of a two-dimensional lattice
\cite{tzeng,wu}. 

On the other hand, in a previous work, two of us used the transfer
matrix approach 
to calculate the entropy of flexible chains with $M$ monomers, that we call
$M$-mers, as 
a function of the density $\rho$ of sites of the lattice occupied by
monomers \cite{wgdstilck}. 
Since only the infinite energy related to
the excluded 
volume interaction is considered, this is an athermal
problem. The entropy was calculated through a numerically exact
procedure for the model defined on stripes of finite widths $n$ and
infinite length, with periodical boundary conditions in the
transverse direction. The sequence of results for stripes of
finite widths was then extrapolated to the two-dimensional limit by means
of finite-size scaling procedures. 

By using computational methods and the asymptotic theory of Pemantle
and Wilson \cite{pw}, 
Kong \cite{k07,kong1,kong2}, obtained, among other results, exact
asymptotic expansions for the entropy of the dimers 
placed on stripes in the high density region. 
It may be appropriate to mention that the free energy $f(\rho)$
defined by Kong is actually the adimensional entropy per lattice site
$s(\rho)=S(\rho)/(Nk_B)$, 
which is related to the Helmholtz free energy of the system through 
\begin{equation}
s(\rho)=-\frac{F(\rho)}{Nk_BT},
\end{equation}
where $k_B$ is the Boltzmann constant and we have set the constant
internal energy of the system equal to zero.  
He observed that the
amplitude of the first term in this expansion  
depends on the parity of the lattice width, $n$,
\begin{eqnarray}
s_n(\rho)\sim s_n(1)-\varphi(1-\rho)\ln(1-\rho),
\label{asympt}
\end{eqnarray}
where
\begin{displaymath}
\varphi=\left\{\begin{array}{ll}
1&\textrm{if n is odd}\\
\frac{1}{2}&\textrm{if n is even.}\\
\end{array}\right.
\end{displaymath}
The asymptotic behavior of the free energy in the low density limit
$\rho \to 0$ was also studied by Kong in the monomer-dimer problem
\cite{k07}.  

It is of interest to consider how the asymptotic results found by Kong
are generalized if chains with more than two monomers are considered.
In this work, using the transfer matrix approach to calculate the entropy
for finite stripes with width $n$, we obtain estimates for the amplitude 
$\varphi$ not only for the dimer case, but also for larger chains
$M=3,4,\ldots,9$. We find that in general expression (\ref{asympt})
describes 
well the behavior of $s(\rho)$ in the high density limit, but the
number of different values of the amplitude, as well as their numerical
values, change as chains with different molecular weights $M$ are
considered. Also, we study the low density limit for the same values
of $M$ mentioned before. 

This paper is organized as follows. In section \ref{calc} we present
the expressions 
used to estimate the amplitudes $\varphi$ and we discuss how the
transfer matrix for the problem is defined and obtained. The numerical
results may be found 
in section \ref{num} and the conclusions are presented in section
\ref{con}. 

\section{Useful expressions, definition of the transfer matrix and its
  construction}
\label{calc}
Although it is rather natural to study the system in the canonical
ensemble, where the number of monomers on the lattice is fixed, to
apply the transfer matrix technique it is convenient to allow this
number to fluctuate.
We thus define the grand-canonical partition function as
\begin{eqnarray}
\Xi(z) = \sum_{p}z^{pM}\Gamma(M,N,p),
\end{eqnarray}
where $z$ is the activity of a monomer and $\Gamma(M,N,p)$ is the
number of ways to place $p$ chains with $M$ monomers each on the
lattice with $N$ sites. The density of monomers may be written
as
\begin{eqnarray}
\rho(z)=z\frac{d}{dz}\phi(z),
\end{eqnarray}
where $\phi(z)$ is the thermodynamic potential per lattice site, defined
as
\begin{eqnarray}
\phi(z)=\lim_{N\rightarrow\infty}\frac{1}{N}\ln\Xi(z).
\end{eqnarray}

In the thermodynamic limit we may use a Legendre transformation
to rewrite the potential as
\begin{eqnarray}
\phi(z)\sim\max_{\rho}\{\rho\ln z+s(\rho)\},
\end{eqnarray}
which implies that
\begin{equation*}
\frac{ds}{d\rho}=-\ln z,
\end{equation*}
and therefore the entropy will be given by
\begin{eqnarray}
\label{fen}
s(\rho)=-\int_{0}^{\rho}\ln z(\rho^\prime)d\rho^\prime,
\end{eqnarray}
with $s(0)=0$.

Now, we may attempt to generalize the high density expansion for
dimers and suppose that the behavior of the entropy in 
this region is given by the expression
\begin{eqnarray}
\label{rela}
\frac{s_n(\rho)-s_n(1)}{(1-\rho)}=A_n-\varphi_n\ln(1-\rho),
\end{eqnarray}
thus, using the equation (\ref{fen}) we obtain
\begin{eqnarray}
\int_{\rho}^{1}\ln z_n(\rho^\prime)d\rho^\prime=A_n(1-\rho)-
\varphi_n(1-\rho)\ln(1-\rho). 
\end{eqnarray}
Deriving this last equation with respect to $\rho$, we have
\begin{eqnarray}
\label{eq}
\ln z_n=C_n-\varphi_n\ln(1-\rho),
\end{eqnarray}
where $C_n=A_n-\varphi_n$. This  expression was useful to obtain
evidences that for all cases we studied the asymptotic behavior
supposed in equation (\ref{rela}) is valid, allowing us to estimate the
amplitudes $\varphi_n$

\subsection{Transfer matrix}

To build the transfer matrix for this problem we 
define a stripe of width $n$ on the square
lattice in the plane $(x,y)$, so that $1\leq x\leq n$ 
and $-\infty\leq y\leq\infty$. The position of a site may be
defined by $(x,y)$, where $1 \leq x \leq n$ and $y$ are integer
numbers. Periodic boundary conditions
are assumed in both directions. Generalizing the prescription
due to Derrida \cite{der} for the transfer matrix of an infinite chain
placed in stripes, we  
define the state of a set of $n$ vertical bonds of the lattice
connecting the sites at $y_0-1$ to the sites at $y_0$ specifying the
number of monomers already connected to this bond located on sites
with $y<y_0$ (in the range
$[0,M-1]$) and the pairs of bonds which are connected to each other by
a chain whose monomers are all located at sites with $y<y_0$. This
last information is essential to prevent the presence of
rings in the allowed configurations.
With the information above about the configuration of the
vertical bonds located between $y_0-1$ and $y_0$, we may find all
possible configurations of the vertical bonds between $y_0$ and
$y_0+1$, thus defining a transfer matrix. Actually, it is not
difficult to develop an algorithm for the steps involved in this
procedure, which allows us to obtain the elements of the transfer
matrix exactly. Restriction in memory and computer time set an upper
limit to the values of $n$ and $M$ we are able to handle, since the
size of the transfer matrix grows very fast as they increase.
More details about this procedure may be found in the
previous paper \cite{wgdstilck}.

Once the transfer matrix $\mathcal{T}$ is obtained, we may find
the entropy in the thermodynamic limit using the largest eigenvalue of
this matrix $\lambda$. The grand-canonical 
partition function is related to the transfer matrix by the expression
\begin{eqnarray}
\Xi(z)=Tr(\mathcal{T}^{\ell}),
\end{eqnarray}
where $\ell$ is the length of the stripe and we adopt periodical
boundary contitions in the longitudinal direction as well.
The density of monomers in the thermodynamic limit $\ell \to
\infty$, $\rho(z)$, will then be given by 
\begin{eqnarray}
\rho(z)=\lim_{N\rightarrow\infty}\frac{z}{N}\frac{d}{dz}\ln\Xi(z)=
\frac{z}{n}\frac{d}{dz}\ln\lambda.
\label{rhoz}  
\end{eqnarray}
Thus, using the expressions (\ref{eq}) and (\ref{rhoz}) and supposing
that the behavior in the  
high density limit of the entropy for chains with $M$ monomers is
given by the relation (\ref{rela}), we may
estimate the amplitudes $\varphi$ for the set the molecular weights
$M$ widths of the stripe $n$ we were able to handle.

\section{Numerical results}
\label{num}

A roughly exponential growth of the size of the transfer matrix with
both the molecular weight $M$ and the width of the stripe $n$ prevents
us from obtaining results for larger chains or stripes. In this paper we show
results for chains with molecular weight ranging between 2 and 9 and widths
of the stripes ranging between 2 and 12 (dimers case) and 2 ($M=9$ case).
First, we analyze the high density limit behavior of the entropy for the
molecular weights and widths of the stripes mentioned. Later, we will turn
our attention to the low density limit for the same cases. For each
pair of values for $n$ and $M$, we obtain the elements of the transfer
matrix using the algorithm mentioned above in a numerically exact
fashion. Then, we take 
advantage of the $\mathbb{C}_n$ symmetry of the states 
and use the power method to find the largest eigenvalue of the matrix
$\lambda$ and its derivative with respect to $z$. 

\subsection{High density limit}

\begin{center}
\begin{figure}[h!]
\includegraphics[scale=0.25]{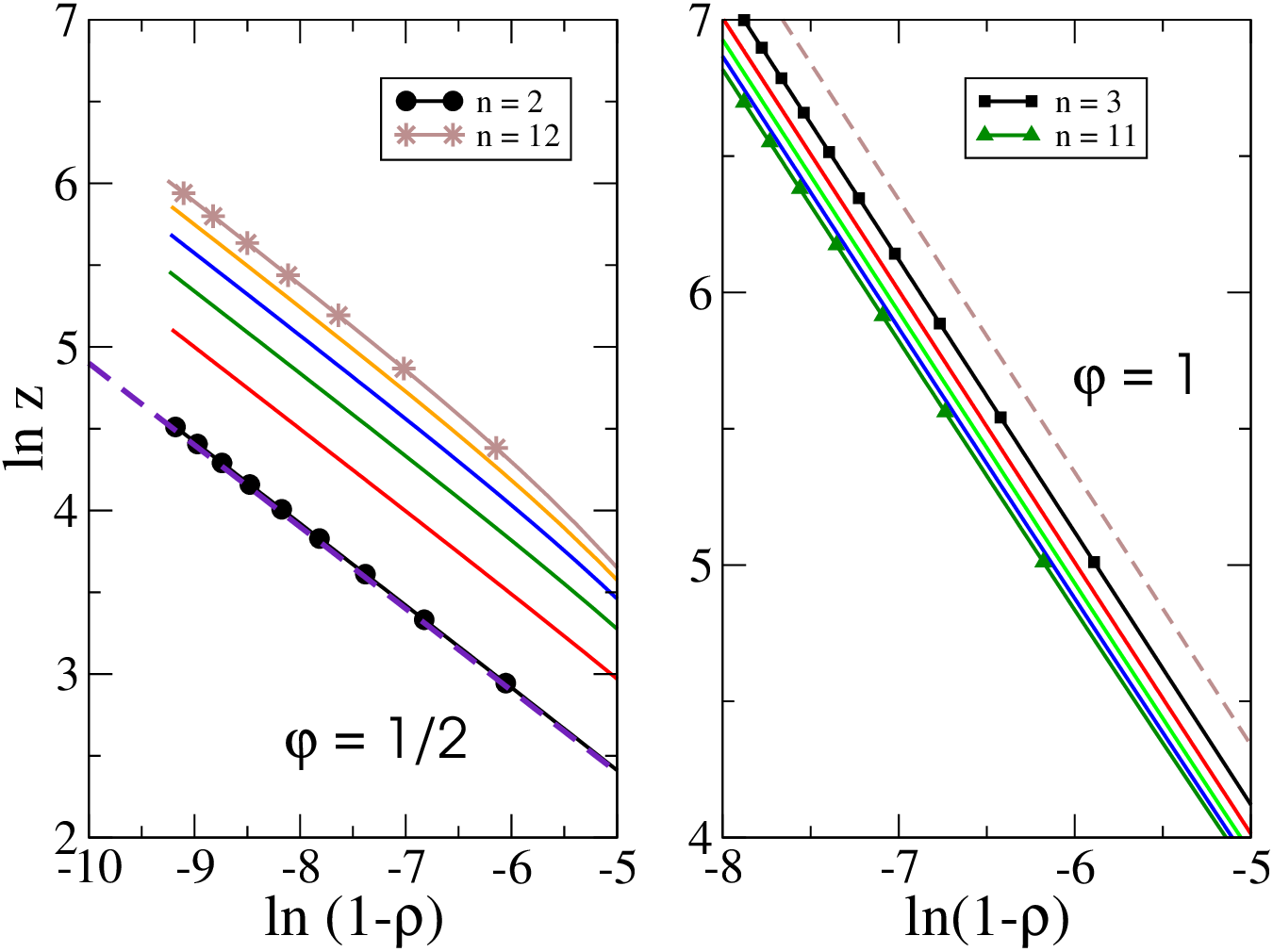}
\caption{Results for the amplitude $\varphi$ for dimers ($M=2$).
At left panel the widths of the stripes are even values ranging
between $n=2-12$, while at right panel results for odd values of the
widths between $n=3-11$ are displayed. The dashed lines indicates 
slopes equal to the known value of $\varphi$ in each case.}
\label{fig1}
\end{figure}
\end{center}

We start with our results for dimers,which are in agreement with the 
values obtained by Kong \cite{kong1,kong2}, with $\varphi=1$, if $n$ is odd
and $\varphi=1/2$, if $n$ is even, as is shown in figure \ref{fig1}.

\begin{center}
\begin{figure}[h!]
\includegraphics[scale=0.31]{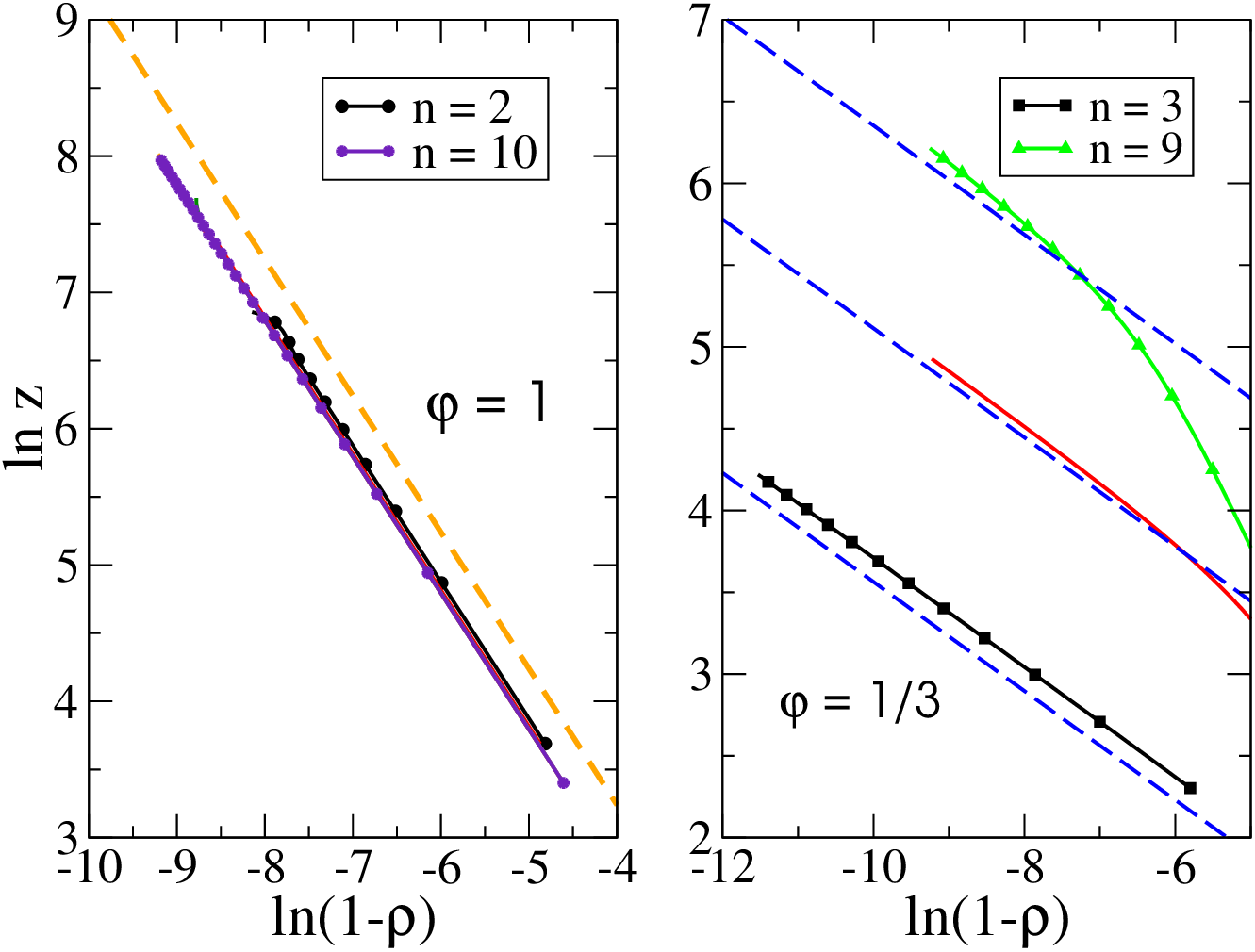}\\
\includegraphics[scale=0.31]{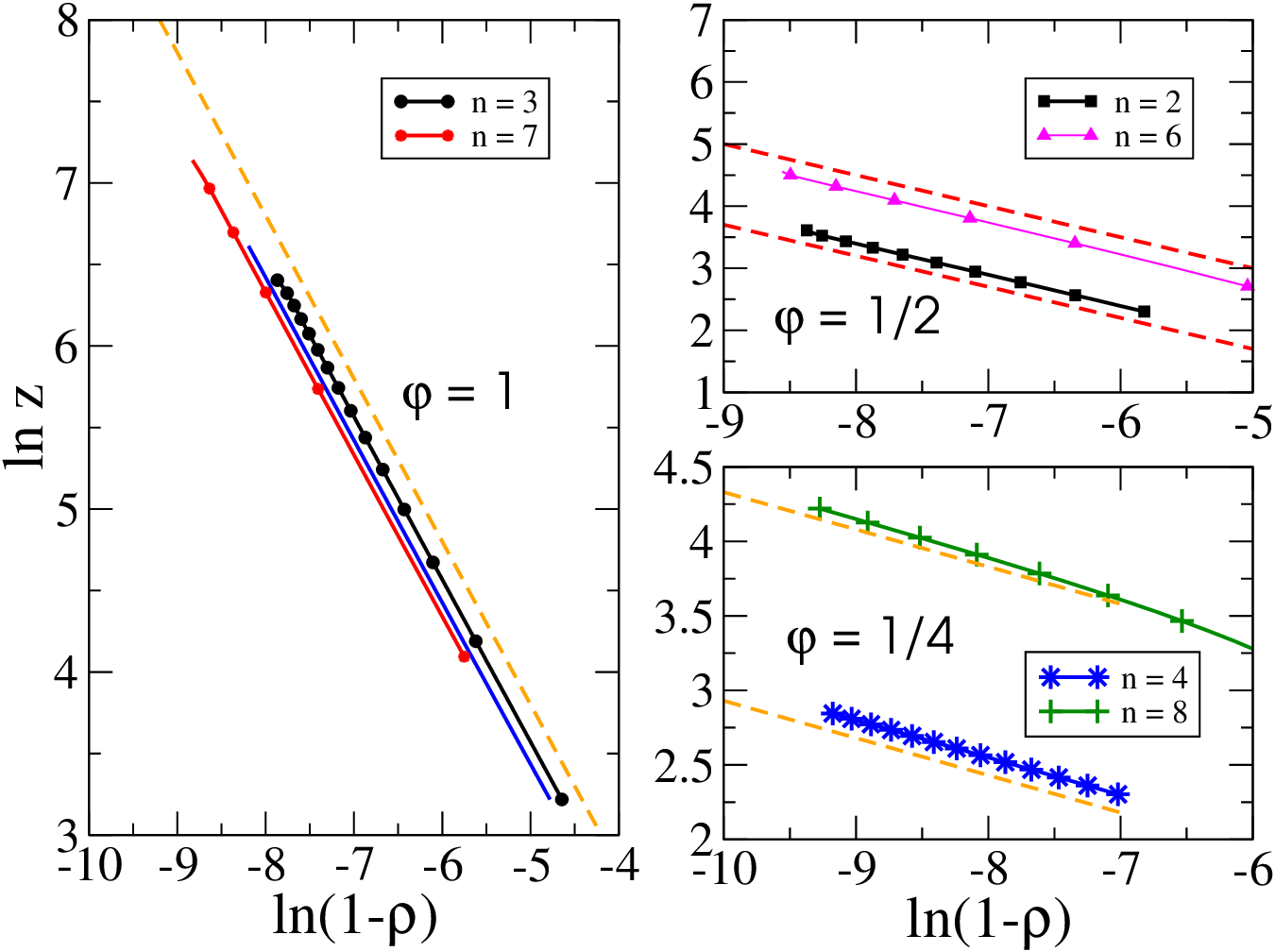}\\
\includegraphics[scale=0.31]{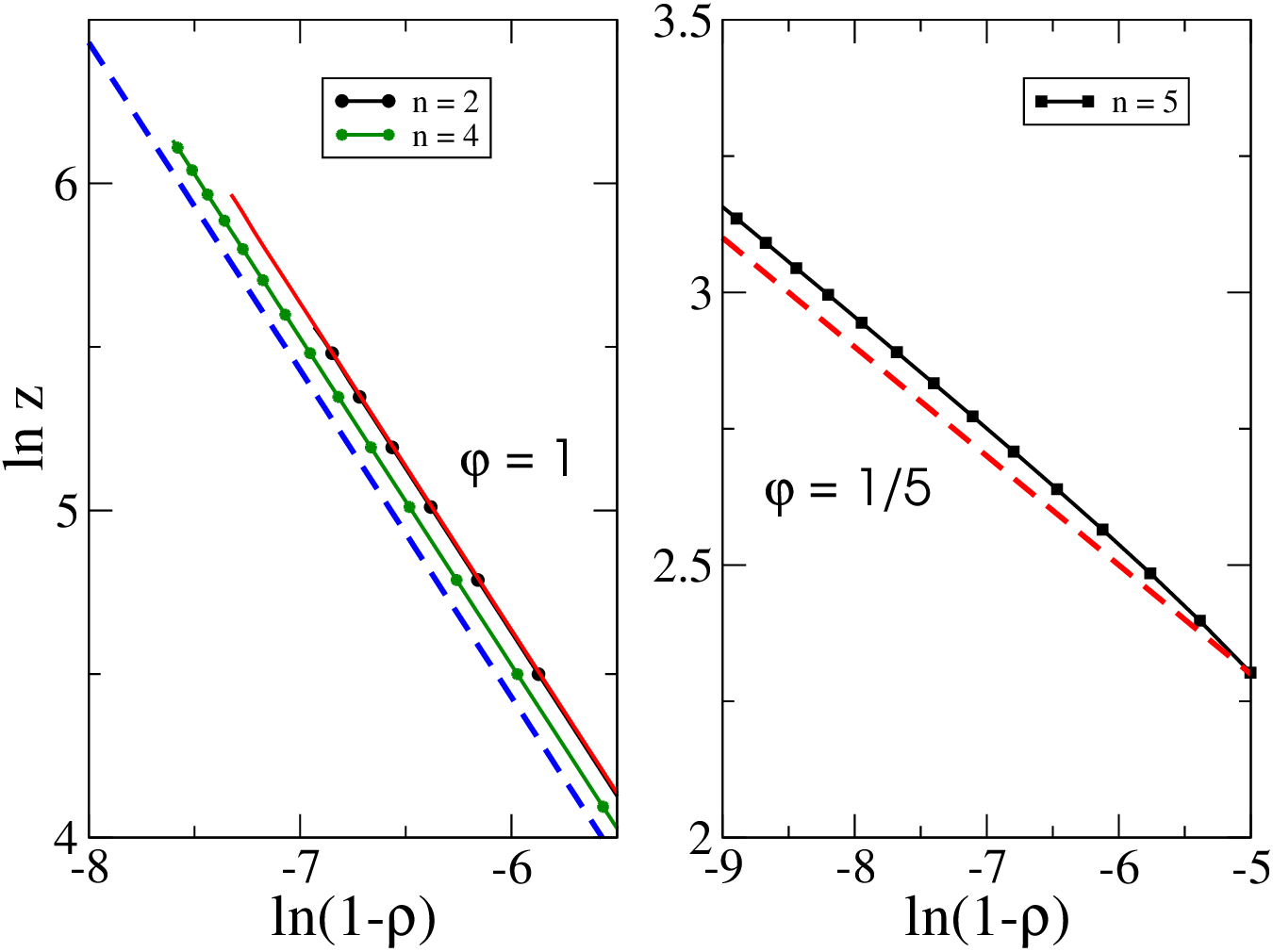}
\caption{Plots which lead to the estimated values of the amplitude
  $\varphi$ for $M=3,4$ and $M=5$. Top: Results for trimers ($M=3$),
  with widths ranging between $n=2$ and $n=10$. In this case for
  widths that are multiples of 3 we find $\varphi=1/3$ for the all
  other widths the results lead to $\varphi=1$. Only the extreme cases
  in each case are labeled.
  Middle: Case $M=4$, for widths in the range $n=2-8$. In this case
  three values of the amplitude are found. For widths which are even
  but not multiples of 4, $\varphi=1/2$. If $n$ is odd, $\varphi=1$
  and $\varphi=1/4$ in all other cases. Bottom: Data for $M=5$, with
  $n=2-5$. Again, values for $\varphi$ are found. If $n$ is a 
  multiple of $5$, then $\varphi=1/5$ and $\varphi=1$ for the other
  cases. The dashed lines are drawn with the conjectured slope
  $\varphi$ in each case.} 
\label{fig2}
\end{figure}
\end{center}

For chains with $M=3$ (trimers), our results also lead to two values of the
amplitude:  
$\varphi=1/3$ for widths that are multiples of 3 and $\varphi=1$
otherwise.
Our results for tetramers $M=4$ show three values for the
amplitude $\varphi$,
\begin{displaymath}
\varphi=\left\{\begin{array}{ll}
1&\textrm{if n is odd}\\
\frac{1}{2}&\textrm{if n is a multiple of 2, but not of 4}\\
\frac{1}{4}&\textrm{otherwise.}
\end{array}\right.
\end{displaymath}
Finally, in the case $M=5$ (pentamers), we found $\varphi=1/5$, if
$n$ is a multiple of 5 and $\varphi=1$, in 
all other cases. Unfortunately, in this case we only were able to
consider widths up to $n=5$, due to computational limitations
caused by the fast growth of the transfer matrix with $n$.
All these results are summarized in figure \ref{fig2}. 

\begin{center}
\begin{figure}[h!]
\includegraphics[scale=0.32]{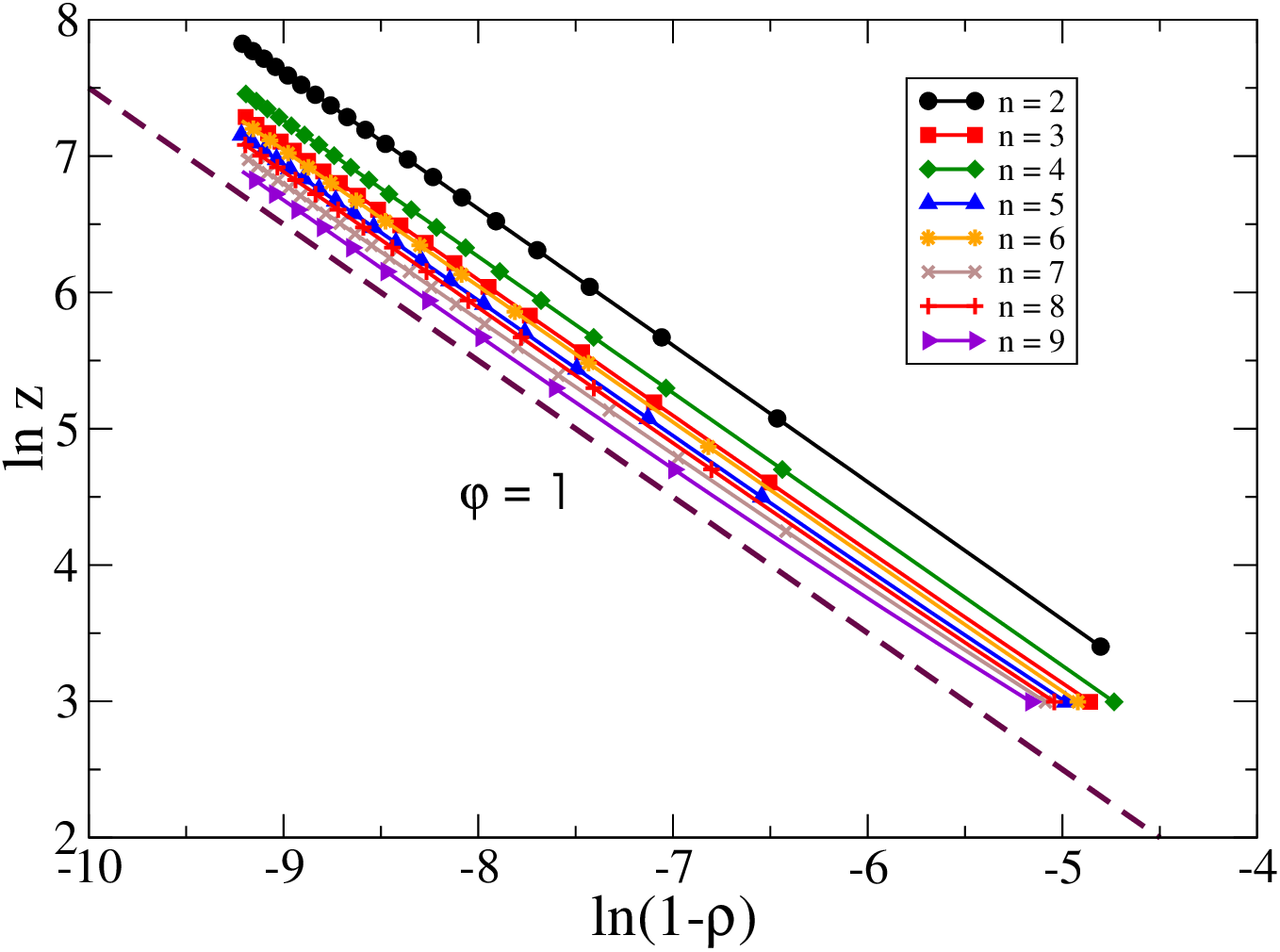}\\
\caption{Polymer case $(M\rightarrow\infty)$, in which only one amplitude
$\varphi=1$ was found for all widths studied ($n=2,3,\ldots,9$).}
\label{fig3}
\end{figure}
\end{center}

In our calculations, numerical errors set an upper limit to the
densities we may consider. As we approach the full occupancy limit
$\rho \to 1$, the values of the activity $z$ become very large and
this makes the numerical errors in the calculations grow. Finally, we extend
our analysis to the polymer case ($M \to \infty$). As was shown
in \cite{wgdstilck}, the entropy for this case shows different finite size
scaling corrections for even and odd widths. Nevertheless, the data
show a single amplitude $\varphi=1$ in this limit, as may be seen in
figure \ref{fig3}. 

We may conclude that the results shown above suggest that the
dependence of the entropy with the width of the stripe
obtained by Kong for the dimer model may be extended
for larger chains, including more possibilities for the
amplitude $\varphi$. Apparently, the values of the amplitude are
related to the
split of the leading eigenvalues of the transfer matrix
into subsets with different finite size scaling behaviors found in
\cite{wgdstilck}.
These splits seem to be related to frustration effects
in the limit of full occupancy. An empirical rule that
we found to label these subsets is the determination of an
integer $\alpha$ which minimizes the relation
\begin{eqnarray}
\label{alprel}
\alpha n=kM,
\end{eqnarray}
where $k$ is the smallest integer for which a solution is found. The
widths associated to the same value 
of $\alpha$ share the same estimate of the amplitude, as may be seen
in table \ref{tab}. Another way to look at relation (\ref{alprel}) is
to rewrite it as $\alpha/k=M/n$, so that $\alpha$ is the numerator of
the fraction $M/n$ after it is simplified. We notice that $\alpha$ may
be interpreted as 
the length of the smallest rectangle of width $n$ which may be totally
filled by chains with $M$ monomers each. The number of chains we can
place in this rectangle is $k$.
We notice that if $M$ is prime, we find $\alpha=M$ for all widths
$n$ which are not multiples of $M$, while $\alpha=1$ if $n$ is a
multiple of $M$. In general, the number of different values for
$\alpha$ is equal to the number of divisors of $M$, including 1 and
$M$ itself.
\begin{table}[h!]
\caption{Values for the integer $\alpha$,
which satisfies the relation (\ref{alprel}) for
some widths and molecular weights.}
\label{tab}
\begin{center}
\begin{tabular}{c c c c c c c c c c c c c c c c c c c}  \hline\hline
$M$& $n$ & $\alpha$ & $k$ & & $M$&$n$&$\alpha$&$k$& &$M$&$n$&$\alpha$&$k$&
&$M$&$n$&$\alpha$&$k$\\ \hline\hline
2&2&1&1& &3&2&3&2& &4&2&2&1& &5&2&5&2\\
 &3&2&3& & &3&1&1& & &3&4&3& & &3&5&3\\
 &4&1&2& & &4&3&4& & &4&1&1& & &4&5&4\\
 &5&2&5& & &5&3&5& & &5&4&5& & &5&1&1\\
 &6&1&3& & &6&1&2& & &6&2&3& & &6&5&6\\
 &7&2&7& & &7&3&7& & &7&4&7& & &7&5&7\\
 &8&1&4& & &8&3&8& & &8&1&2& & &8&5&8\\
 &9&2&9& & &9&1&3& & &9&4&9& & &9&5&9\\
\vdots&\vdots&\vdots&\vdots& &\vdots&\vdots&\vdots&\vdots&
&\vdots&\vdots&\vdots&\vdots& &\vdots&\vdots&\vdots&\vdots \\
6&2&3&1& &7&2&7&2& &8&2&4&1& &9&2&9&2\\
 &3&2&1& & &3&7&3& & &3&8&3& & &3&3&1\\
 &4&3&2& & &4&7&4& & &4&2&1& & &4&9&4\\
 &5&6&5& & &5&7&5& & &5&8&5& & &5&9&5\\
 &6&1&1& & &6&7&6& & &6&4&3& & &6&3&2\\
 &7&6&7& & &7&1&1& & &7&8&7& & &7&9&7\\
 &8&3&4& & &8&7&8& & &8&1&1& & &8&9&8\\
 &9&2&3& & &9&7&9& & &9&8&7& & &9&1&1\\
\vdots&\vdots&\vdots&\vdots& &\vdots&\vdots&\vdots&\vdots&
&\vdots&\vdots&\vdots&\vdots& &\vdots&\vdots&\vdots&\vdots \\
\\
\hline\hline
\end{tabular}
\end{center}
\end{table}

Our results suggest that the amplitude $\varphi$ is related
to the integer $\alpha$ by the simple relation,
\begin{eqnarray}
\varphi=\frac{\alpha}{M},
\end{eqnarray}
which is obeyed by all numerical results presented here.
However, we have no physical or mathematical argument to justify
this hypothesis. Nevertheless, the numerical evidences 
support this relation. If we suppose that the polymer limit $M \to
\infty$ is approached by a sequence of prime values of $M$, the we
would conclude that $\alpha=M$ in this limit, which is consistent with
the observed amplitude $\varphi=1$.

\begin{center}
\begin{figure}[h!]
\includegraphics[scale=0.34]{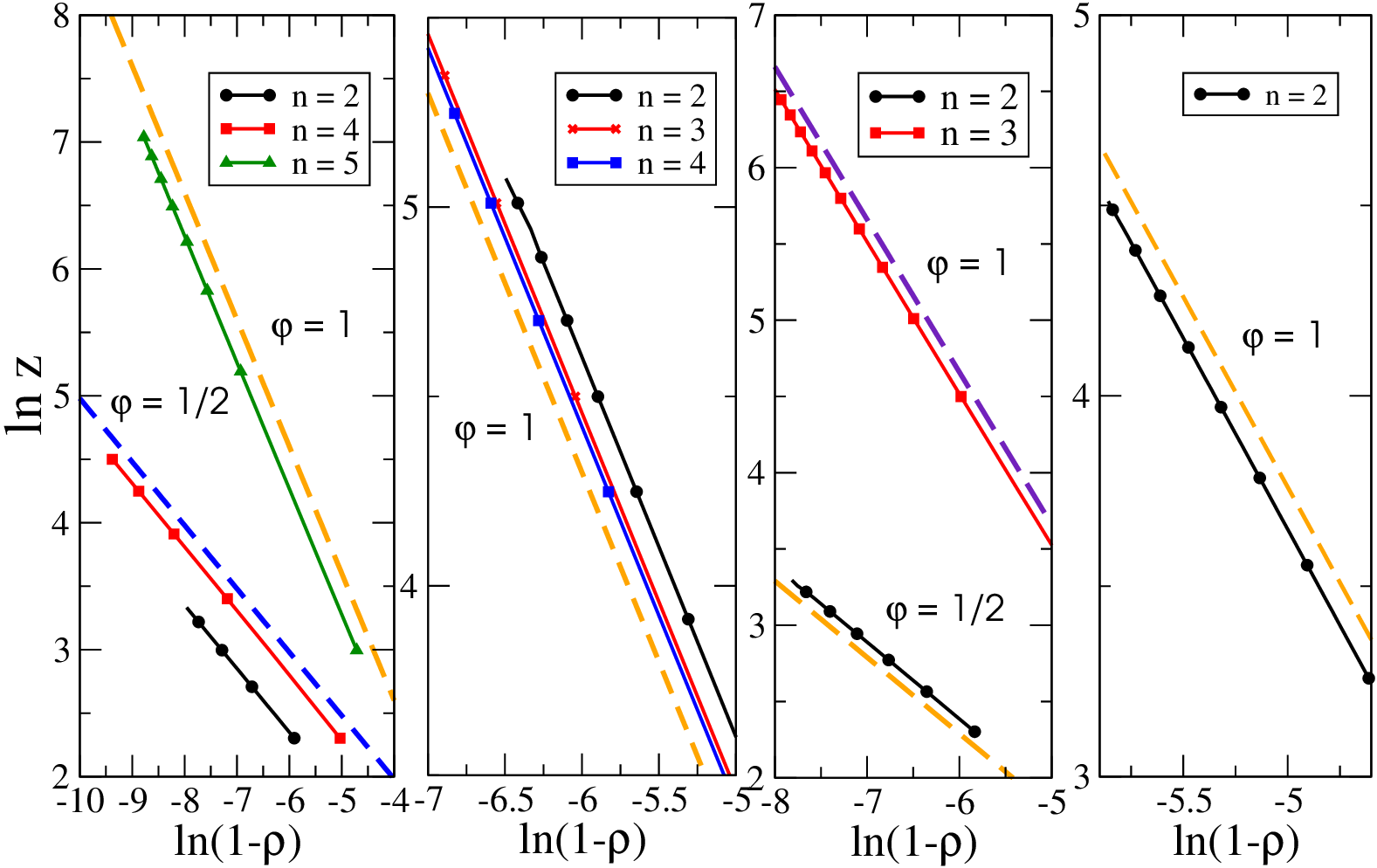}
\caption{Results for the entropy in the high density limit for chains
  with molecular weight $M$ between 6 and 9, in increasing values from
  left to right. The observed values of the amplitudes are consistent
  with the conjecture made in the text.} 
\label{fig4}
\end{figure}
\end{center}

In order to test the conjecture above for the amplitude of the high
density limit asymptotic form of the entropy, we extended the transfer
matrix calculations to higher values of the molecular weight $M$,
although with increasing values of $M$ we are restricted to decreasing
maximum widths $n$. As may be seen in figure \ref{fig4}, all results
for chains with molecular weights $M=6,7,8,9$ are consistent with the
conjecture for the amplitudes $\varphi$. There are cases, like
$M=6$, $n=3$, where numerical errors prevented us from obtaining results
at densities high enough to observe the asymptotic behavior. An
example where the asymptotic regime also was reached only at very high
densities may be seen in results for $M=3$, $n=9$ in figure
\ref{fig2}. As a consequence of these limitations, only some of the
possible values for the amplitude $\varphi$ are observed in the
results for chains with larger molecular weight presented in figure
\ref{fig4}. 

\subsection{Low density limit}

In the low density limit $\rho\rightarrow 0$, Kong \cite{kong1,kong2}
obtained 
from an asymptotic expansion the following behavior for the entropy
of dimers placed in a stripe with width $n$,
\begin{eqnarray}
s_n(\rho)\sim -\frac{\rho}{2}\ln\rho+\mathcal{O}(\rho).
\end{eqnarray}
The expression above resembles the one predicted by the mean-field
approximation \cite{flory,mj}, which is
\begin{eqnarray}
s(\rho)&=&-(1-\rho)\ln(1-\rho)-\frac{\rho}{M}\ln\left(\frac{2\rho}{M}\right)
+\nonumber\\
&+&\left(1-\frac{1}{M}\right)(\ln q-1),
\end{eqnarray}
where $q$ is the coordination number. For small values of the density
$\rho$, the
leading contribution comes from the second term, which for dimers is
identical to the one obtained by Kong.

It is then rather natural to conjecture that for other values of $M$
the leading contribution to the entropy is the one predicted by the
mean field approximation, since for small densities the interchain
interactions may be neglected. Thus, we conjecture that for all values
of the molecular weights in the low density limit we have the
asymptotic behavior
\begin{eqnarray}
s(\rho)\sim\frac{\rho}{M}\ln\left(\frac{2\rho}{M}\right).
\label{ldb}
\end{eqnarray}

\begin{center}
\begin{figure}[h!]
\includegraphics[scale=0.28]{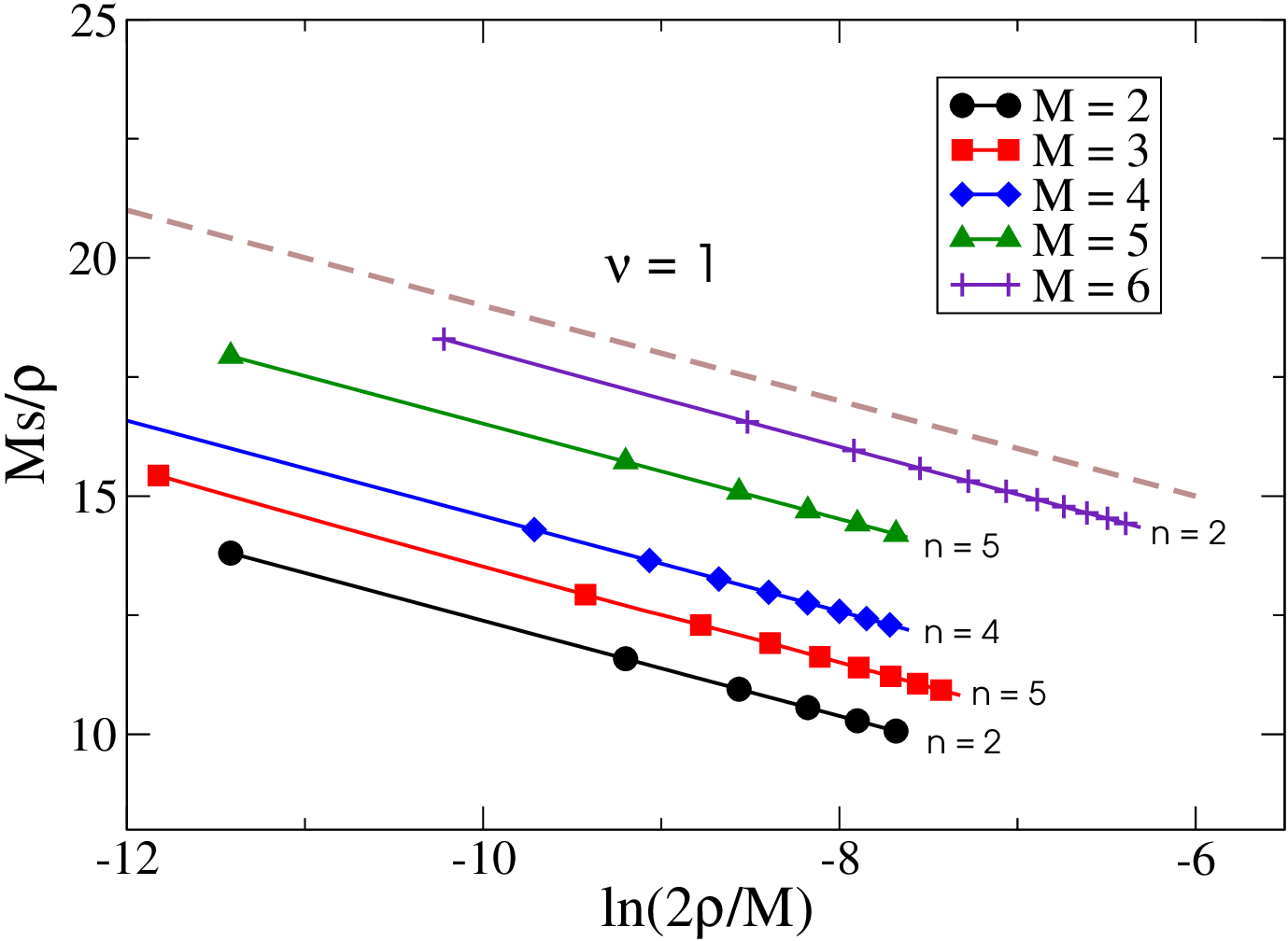}
\caption{Results for the low density limit with $M$ varying between
  $M=2$ and $M=6$. In all cases, independently of the
  width analyzed, we observe linear relations with slope $\nu=1$
  (indicated by the dashed line),
  satisfying the mean-field prediction.} 
\label{fig5}
\end{figure}
\end{center}

In order to test expression (\ref{ldb}) in the low density limit,  
we built curves for different molecular weights and widths of stripes.
Our results, as shown in the figure \ref{fig5}, are consistent with
the conjecture
for all the chains analyzed here, since curves of $Ms(\rho)/\rho$ as a
function of $\ln(2\rho/M)$ are linear with a slope equal to 1 in
all cases.

\section{Conclusions}
\label{con}
In this work, we study the asymptotic behavior of the entropy for 
chains placed on stripes using the transfer matrix
approach. Generalizing the results by Kong \cite{k07,kong1,kong2} for the
case of dimers in the high density limit, we conjecture similar
asymptotic forms for the entropy 
of chains with larger molecular weights. We propose an empirical rule
to determine 
the number of different amplitudes of the asymptotic behavior close to
the full occupancy limit, as well as to find the values of these
amplitudes. 
Unfortunately, computational limitations prevent us from studying larger
chains and larger widths. However, it seems reasonable expect that our
results apply also to other cases. An analytic approach
similar to the one made by Kong \cite{kong1,kong2}, using the Pemantle-Wilson
asymptotic 
theory \cite{pw} would be very useful to support our
conclusions. However, it may not be easy
to carry out this task for chains with molecular weight
larger than $M=2$.

Another interesting result is that in the low density limit the free
energy is well fitted by the mean-field approximation \cite{flory,mj},
for any width $n$ for all the values of $M$ we were able to
consider. Again, it would be interesting to obtain an exact result 
in this limit in order to verify this conclusion.

\section*{Acknowledgment}
W.G. Dantas acknowledges the financial
support from  Funda\c{c}\~ao de Amparo \`a Pesquisa do
Estado de S\~ao Paulo (FAPESP) under Grant No. 05/04459-1 and
JFS acknowledges funding by project  PRONEX-CNPq-FAPERJ/171.168-2003.


\begin{thebibliography}{99}
\bibitem{fowler37}R.H. Fowler and G.S. Rushbrooke, \emph{Trans. Faraday Soc.}
\textbf{33},1272 (1937).
\bibitem{kast2}P.W. Kasteleyn, \emph{J. Math. Phys.}
\textbf{4},287 (1963).
\bibitem{fis2}M.E. Fisher, \emph{J. Math. Phys.}
\textbf{7},1776 (1966).
\bibitem{fan}C. Fan and F. Wu, \emph{Phys. Rev. B}
\textbf{2},723 (1970).
\bibitem{sn74}S. R. Salinas and J. F. Nagle, \emph{Phys. Rev. B}
  \textbf{11}, 4920 (1974).
\bibitem{nagle}J.F. Nagle, C.S.O. Yokoi and S.M. Battacharjee, in
  \emph{Phase Transitions 
and Critical Phenomena}, edited by C. Domb and J. Lebowitz (Academic
  Press, New York, 1989), Vol. 13.
\bibitem{fis}M.E. Fisher, \emph{Phys. Rev.}
\textbf{124},1664 (1961).
\bibitem{kast}P.W. Kasteleyn, \emph{Physica}
\textbf{27},1209 (1961).
\bibitem{temp}H.N.V. Temperley and M.E. Fisher, \emph{Philos. Mag.}
\textbf{6},1061 (1961).
\bibitem{tzeng}W.J. Tzeng and F.Y. Wu, \emph{J. Stat. Phys.}
\textbf{110},671 (2003).
\bibitem{wu}F.Y. Wu, \emph{Phys. Rev. E}
\textbf{74},020104 (2006); \textbf{74},039907(E) (2006).
\bibitem{wgdstilck}W.G. Dantas and J.F. Stilck, \emph{Phys. Rev. E}
\textbf{67},031803 (2003).
\bibitem{pw}R. Pemantle and M.C. Wilson, \emph{J. Comb. Theory Ser. A}
\textbf{97}, 129 (2002).
\bibitem{k07}Y. Kong, \emph{Phys. Rev. E} \textbf{75}, 051123 (2007).
\bibitem{kong1}Y. Kong, \emph{Phys. Rev. E}
\textbf{74}, 011102 (2006).
\bibitem{kong2}Y. Kong, \emph{Phys. Rev. E}
\textbf{74}, 061102 (2006).
\bibitem{der}B. Derrida, \emph{J. Phys. A}
\textbf{14}, L5 (1981).
\bibitem{flory}P.J. Flory, \emph{Principles of Polymer Chemistry} (Cornell
Uninversity Press, Ithaca, 1953).
\bibitem{mj} J. F. Stilck and M. J. de Oliveira,
\emph{Phys. Rev. A} \textbf{42}, 5955 (1990).

\end{thebibliography}
\end{document}